\documentstyle[12pt]{article}
\def\ba{\begin{eqnarray}}
\def\ea{\end{eqnarray}}

\def\lb{\label}
\def\be{\begin{equation}}
\def\ee{\end{equation}}

\begin{document}
\title{Geometric transitions, double scaling limits and gauge theories}
\author{Osvaldo P. Santillan \thanks{Departamento de Matematica, FCEyN, Universidad de Buenos Aires, Buenos Aires, Argentina
firenzecita@hotmail.com and osantil@dm.uba.ar}}
\date {}
\maketitle
\begin{abstract}
In the present work certain features of the Penner model, such its enumerative meaning and its relation to the Chern-Simons theory on the 3-sphere, are reviewed. Also, some features related to geometric transitions at the level of the observables are discussed. In this setup, the non commutative five dimensional $U(1)$ Nekrasov partition function is interpreted as a limiting case of a mean value of the Ooguri-Vafa operator for a Chern-Simons model, which is dual to the A-model on a toric Calabi-Yau with $h_{1,1}=1$ and $h_{2,1}=2$. We work out a B-model interpretation of this identification explicitly by considering this model on the corresponding mirror geometries.
\end{abstract}

\tableofcontents

\section{Introduction}
The study of enumerative problems and topological invariants is a topic of major interest in pure mathematics. The study of non perturbative phenomena of gauge theories is a topic of major interest in physics. In the last thirty years there is growing evidence that each of these issues contributes significantly to the development of the other. An example is the Clemmens conjecture \cite{Clemmens} in enumerative geometry which states that there are a finite number of rational curves of a given degree in the quintic Calabi-Yau. This conjecture was studied in the context of topological sigma models and, by use of duality and mirror symmetry arguments, a list of these numbers was presented in \cite{Candelas}. Although the conjecture it is known to be true for curves of lower degree, a proof of this conjecture is not known. But is remarkable that the lower degree number are exactly reproduced by the calculations of \cite{Candelas}, which additionally predicts the numbers for curves of higher degree. The generalization of this calculation for more general Calabi-Yau geometries is one of the roots of the Gromov-Witten invariants. These invariants usually take rational values and therefore it is difficult to give them an enumerative meaning. This difficulty was interpreted as the effect of bubbling for holomorphic maps. A partial answer to this problem was given by Gopakumar and Vafa \cite{GV} who find a set of integer numbers which can be constructed in terms of the Gromov-Witten invariants and which may count the number of these holomorphic maps. This conjecture is true for some cases but, to the best of our knowledge, a complete proof is not known.

Another important tool in the study of topological sigma models are geometric transitions, which connect deformed geometries with resolved ones. The classic examples are the two resolutions of the Calabi-Yau conifold, the deformed one is $T^{\ast}S^3$ and the resolved one, which is the fibration $X^{t}={\cal O}(-1)\oplus {\cal O}(-1)\rightarrow P^1$. In general, the geometries of the form $T^{\ast}M$ with $M$ a suitable 3-manifold are Calabi-Yau. When a IIA string theory with a worldsheet with boundaries is defined over such geometries and N branes are placed at the boundaries, the theory admits a dual description in terms of a CS theory with gauge group $SU(N)$ over the base 3-manifold $M$. By summing up all the holes the branes disappear and a dual closed string version is obtained. The main conjecture of \cite{GV} is that this string is defined over the resolved geometry. For the deformed conifold, the background should be $X^{t}={\cal O}(-1)\oplus {\cal O}(-1)\rightarrow P^1$. It is believed that such dualities are true beyond the $S^3$ example.

A privileged link  between algebraic geometry and physics is played by the matrix models. Examples are the Penner \cite{Penner} and the Kontsevich ones \cite{Kontsevich}. The Penner one is a powerful tool for calculating the Euler characteristics of the moduli space of Riemann surfaces of genus $g$ and $s$ punctures. The Kontsevich model allows to calculate the intersection number of the Atiyah-Bott-Morita-Miller-Mumford \cite{Mumford}-\cite{Miller} classes of these moduli spaces and is equivalent to two dimensional topological gravity \cite{Witold}. At the present there are known several topological field theories admitting a matrix description. An example are the CS theories mentioned above, and also IIA and IIB strings models on local Calabi-Yau geometries.

Topological field theories, matrix models and dualities are key ingredients for geometrical engineering of supersymmetric gauge theories. These methods
were shown to be successful for reproducing the Seiberg-Witten anzatz \cite{SW} for the Wilsonian effective action $N=2$ super Yang-Mills theory with gauge group $SU(2)$, and were further applied to other gauge theories with matter. Nevertheless, a complete proof of the validity of this anzatz was obscured by the fact that the instanton measure becomes very hard to deal with when the instanton charge grows. A notable advance was done by Nekrasov \cite{Nekrasov} who employed localization techniques in calculating the ADHM instanton integration. These approach itself is related to cohomological topological field theories. The outcome of this investigation is the so called Nekrasov partition function, which reduce the instanton calculation of Seiberg-Witten to a
combinatorial expression in terms of summations over Young diagrams \cite{Nekrasov}-\cite{Losev}. The result is in agreement with the Seiberg-Witten anzatz. Also it was obtained in \cite{Nekrasov} a one parameter deformation of this partition in terms of equivariant instanton calculus of a U(1)$\times$U(1) subgroup of the rotation subgroup of SO(4)$\simeq$ SU(2)$\times$ SU(2) acting on Euclidean space-time $R^4$. This partition has two parameters
$\epsilon_1$, $\epsilon_2$ instead of one $g_s$ which are interpreted as the angular velocities for the U(1)$\times$U(1) subgroup of rotation. The  deformed version $Z(\mu, \epsilon_1, \epsilon_2)$ broke supersymmetries unless the parameters satisfy
$\epsilon_1-=\epsilon_2=g_s$. Additionally the Nekrasov theory has been found to be related to the quantization of algebraic integrable systems \cite{sapa1}-\cite{sapa7}.

In the last year a correspondence which relate the Nekrasov partition function of N = 2 conformal
invariant SU(2) gauge theory and the Virasoro conformal blocks of chiral Liouville field theory with background charge was proposed by Alday-Gaiotto and Tachikawa (AGT) \cite{AGT}. This correspondence is rather interesting, as involve two theories defined in different space time dimensions. Its roots rely in the M5-brane construction of N = 2 gauge theories which was developed in \cite{Gaiott} and further studied in \cite{AGT2}-\cite{AGT5}.  In this construction,
a four-dimensional N = 2 superconformal gauge theory arises from compactification of M5-branes
on a Riemann surface with punctures. The Nekrasov partition
function of a N = 2 superconformal gauge theory turns to be related to the Liouville conformal block for the
Riemann surface. The choice of the background charge of the Liouville theory determine the parameters $\epsilon_i$ of the Nekrasov deformation. A non trivial consistency check for the AGT conjecture for the case $N_f=4$ of the conjecture was reported in \cite{Giribet}-\cite{Giribet5}.

Recently, a novel point of view concerning the AGT conjecture was introduced by Dijkgraaf and Vafa \cite{DV} (see also \cite{DV2}), who proposed
a Penner like matrix model whose classical spectral curve reproduces the Gaiotto curve \cite{Gaiott}. The last is known to reduce in certain limit to the usual Seiberg-Witten curve. The main conjecture is that the partition of DV matrix model gives Virasoro conformal blocks of chiral Liouville field theory and simultaneously gives the undeformed Nektrasov partition function, thus providing a link between both theories. This line of though was push forward in \cite{gimme}-\cite{gimme4} and relations with generalized Selberg integrals can be found in \cite{Iguri}-\cite{Iguri2}. Additionally it was presented in \cite{DV} a matrix model which corresponds to the deformed Nekrasov partition function $Z(\mu, \epsilon_1, \epsilon_2)$.

The DV realization of the AGT conjecture is specially interesting, as it implies a revival for methods of enumerative geometry for studying problems in gauge theories. This realization involve several sophisticated methods of modern mathematical physics such as large N dualities for topological strings and enumerative invariants \cite{GV}-\cite{Ooguvafa2}, CS theories and matrix models \cite{Aganagic}-\cite{Agavafa2}, geometric transitions \cite{Grassi}, Calabi-Yau crystals \cite{Crystal} mirror symmetry and maps between matrix models and Liouville theory \cite{mapeo}. These features makes that formulation considerably attractive.

The aim of the present work is to discuss an small but still rich realization of the ideas described in the previous two paragraphs, and is organized as follows. In section 2 we review several features of the Penner model. We emphasize its enumerative meaning for calculating the Euler characteristic of the moduli space $M_g^s$ of complex structures for Riemann surfaces of genus $g$ and $s$ punctures. We show that
its double scaling limit gives direct information about the perturbative and non perturbative part of the topological CS theory with gauge group $SU(N)$ over $S^3$. In section 3 we review geometrical transitions in the A-model at the level of the partition function and the observables. We also interpret the N=2 abelian Nekrasov partition function as certain limiting case of the Ooguri-Vafa operator corresponding to a CS theory on certain toric geometry with $h_{1,1}=1$ and $h_{2,1}=2$. In section 4 we elaborate the same interpretation in the B-model setup, by considering corresponding mirror Calabi-Yau geometries.

\section{Matrix models, Chern-Simons theories and double scaling limits}
\subsection{The Penner model}

A classic example of a matrix model with logarithmic interaction is the Penner model \cite{Penner}, which compute the orbifold Euler characteristic $\chi(M_g^s)$ of the moduli space $M_g^s$ of complex structures on a Riemann surfaces of genus $g$ with $s$ punctures (or marked points). This correspondence between the topological invariants of $M_g^s$ and matrix models is related to an specific set of graphs, which appears when considering a simplicial decomposition of $M_g^s$ for computing the Euler characteristics and which are Feynman diagramms of an specific matrix model, the Penner one.

Any two dimensional metric $g_2$ defined over a Riemann surface is locally conformally flat. This means that the distance element can be parameterized as \be\lb{eres}g_2=e^{\phi}dz\otimes d\overline{z},\ee with $(z, \overline{z})$ a complex coordinate system and $\phi(z,\overline{z})$ an scalar function. Two complex coordinate complex coordinate systems $(z, \overline{z})$ and $(z', \overline{z}')$ are considered equivalent is they are related by a transformation of the form $z'=f(z)$. The moduli space $M_g^s$ parameterize the set of \emph{inequivalent} complex coordinate systems for a Riemann surface of genus $g$ and $s$ marked points. It has dimension $3g-3+s$, and this number is always required to be positive.

The space $M_g^s$ is not smooth, and to calculate its topological invariants is of special interest. One of these invariants, the Euler characteristic, requires a simplicial decomposition. An useful one is provided in terms of trajectories of
Jenkins-Streibel quadratic differentials \cite{Strebel}. Given a Riemann surface with marked points,
there always exist a quadratic differential $\eta=\eta_{zz}(z)dz^2$ with poles at the positions of the punctures. The complex coordinate $u$ defined by $du=\sqrt{\eta(z)}dz$ induces a notion of the length of an arbitrary curve $\gamma$ on the Riemann surface by the Euclidean formula
\be\lb{stre}
l(\gamma)=\int_{\gamma}\mid du\mid.
\ee
The curves for which $\eta$ is positive and real along them are called horizontal and for the negative real case the curves are known as vertical. Horizontal curves describe regular flows along the Riemann surface, except at the positions where the zeros of $\eta$ are located. When approaching a zero of order $n$ with a fixed angle it follows that $\eta\simeq z^n dz^2=e^{i(n+2)\theta}d\rho^2$ and therefore there are $n+2$ angles at which the differential is positive.
This may be paraphrased that at a zero of order $n$ there are $n+2$ horizontal curves meeting. Instead near a pole of order two, the function $\eta$ take negative values. By moving around in the angular direction it follows that $\eta\sim -dz^2/z^2=d\theta^2$ is positive, thus the curves surrounding these points are all horizontal, and form concentric loops.

The set of differentials with the properties described above, and with the further property that the curves that do not pass through zero are closed, do in fact exist. Once a complex structure are fixed they are unique, and therefore characterize the moduli space $M_g^s$ \cite{Strebel}. Any of these differentials determines a metric by (\ref{stre}) and thus notion of length for the edges of the graph and conversely, given a graph with edges of a given length, one can construct the corresponding quadratic differential. The local element (\ref{eres}) shows that a fixed complex structure of a given Riemann surface do not change by an overall scaling. Thus, by varying the lengths of the edges E while keeping the total length L fixed, it is possible to parameterize the space of these differentials and, indirectly, the moduli $M_g^s$.\footnote{More precisely, this cell decomposition parameterize $M_g^s \times R^s_+$, with $R^s_+$ is the octant parameterized by the perimeter $p_i$ of the i-th face. The space $M_g^s \times R^s_+$ is the Mumford decorated space \cite{Decorado}, whose Euler characteristic is known to be equal, up to a sign, to the Euler characteristic of $M_g^s$.} Details of these non trivial assertions can be found for instance in \cite{Lando}.

The horizontal flows for the differentials described above can be though as a graph  with $n+2$ incident edges E in which the zeroes or branch points are vertices. The poles of order two give rise to loops, therefore the number of these loops is equal to the number $s$ of punctures of the Riemann surface. By varying the lengths of the edges, a simplicial decomposition of the moduli space is obtained. Any of these graphs determines a two dimensional surface $F(G)$ for which $G\subset F(G)$ in such a way the inclusion is an homotopy equivalence. This is achieved by decomposing the lines into two ones, giving rise to a ribbon graph, or a fat graph. The two invariants characterizing the surface $F(G)$ are its Euler characteristic $\chi(G)$ and its genus $g(G)$, these invariants can be expressed in terms of the number of boundary components $s(G)$ and the number of k-valent vertices $v_k(G)$ of $G$ by the relations
\be\lb{topin}
\chi(F(G))=\chi(G)=\frac{1}{2}\sum_k (2-k)v_k(G),
\ee
\be\lb{topon2}
g(G)=\frac{1}{4}\bigg(4-2s(G)+\sum_k (k-2)v_k(G)\bigg).
\ee
The Euler characteristic of $M_g^s$ is calculated in terms of this simplicial decomposition, which is equal to the Euler characteristic of the fat graph complex. It is given by \cite{Penner}
\be\lb{orbo}
\chi(M_g^s)=\chi(C^{\ast})=\sum_{G \in\widetilde{G}} \frac{(-1)^{v(G)}}{|\Gamma(G)|},
\ee
with $v(G)$ the number of vertices of the graph $G$. In these terms, the generating function for the orbifold Euler characteristic (\ref{orbo}) is defined by
\be\lb{genfun}
\Phi_c(I, N)=\sum_{-2\chi(G)=I}\frac{(-1)^{\sum_k v_k(G)}N^{s(G)}}{|\Gamma(G)|},
\ee
with the sum is understood over \emph{connected} graphs. The analogous quantity without this restriction is usually
denoted as $\Phi(I, N)$. The function (\ref{genfun}) can be expressed as well as
\be\lb{genfun2}
\Phi_c(I, N)=\sum_{g\geq 0, s\geq 1}\chi(M_g^s)N^s,
\ee
with the sum is done taking into account the constraint $I=4g-4+2s$. Additionally, it is customary to introduce the quantity $Z(N, t)$
\be\lb{partu}
Z(t, N)=\sum_{i\geq 0}\Phi(N, I)(i\sqrt{t})^I,
\ee
which is a polynomial expression in the indeterminate $t$, or the quantity $F(N,t)$ \be\lb{partu2}F(N,t)=\sum_{i\geq 1}\Phi_c(N, 2I)(-t)^I,\ee
for connected graphs.

The quantities (\ref{partu})-(\ref{partu2}) defined in the previous paragraph are the fundamental ones for the Penner theory. From the other side, functions such as $Z(t, N)$ or $F(t, N)$ arise typically in $N\times N$ hermitian matrix models perturbative calculations such as those in \cite{Itzykson}. The partition function of those models is generically
\be\lb{gines}
Z=\int dM(H)\;e^{-N\mathrm{Tr}V(M)}
\ee
in which $M$ is an $N\times N$ hermitian matrix, $V(M)$ is a generic potential and the measure the space $H^{N\times N}$ of hermitian $N\times N$ matrices is given by
$$dM(H)=\prod^N_{i=1}dM_{ij}\prod_{i<j} ({\mathrm{dRe}} M_{ij})({\mathrm{dIm}} M_{ij}).$$
For an arbitrary polynomial potential
$$V(x)=\frac{x^2}{2}+\sum_{k=3}^{\infty}\frac{g_k}{k} x^k,$$
it is known that the quadratic interaction corresponds to a free field and its expansion give the propagator
$$
<M_{ij}M_{kl}>=\int dM(H)\;M_{ij}M_{kl}\;e^{-N{\mathrm{Tr}}V(M)}=\frac{1}{N}\delta_{ik}\delta_{jl},
$$
 and that $k$-powers give rise to $k$-valence vertex. The corresponding diagrams will contain high powers of the coupling constants $g_k$. In fact the partition function of the model can be written as
$$
Z=<e^{N\sum_{k=3}^{\infty}\frac{g_k}{k}{\mathrm{Tr}} M^k}>=\sum_{n_1, n_2,..\geq 0}\prod_{k\geq 1}\frac{(N g_k)^{n_k}}{k n_k!}<\prod_{i\geq 1}({\mathrm{Tr}}M_i)^{n_i}>
$$
$$
=\sum_{n_1, n_2,..\geq 0}\prod_{k\geq 1}\frac{(N g_k)^{n_k}}{k n_k!}\sum N^{-E} N^{F},
$$
where the last sum is related to all the labeled fat graphs with $n_i$ valent vertices and $E$ the number of vertices of a graph and $F$ the number of faces. The summation of the possible labeling of an unlabeled graph modify the symmetry factors, leaving with an expression of the form \cite{difrancesco}
\be\lb{discofever}
Z=\sum_{G} \frac{N^{V(G)-E(G)+F(G)}}{|\Gamma(G)|}\prod_{k\geq 1}g_k^{n_k(G)},
\ee
with the sum is over all the fat graphs $G$, $n_k(G)$ is the number of $k$-valent vertices of the graph $G$ and $V-E+F$ is equal to the Euler characteristic $\chi(G)$ of the graph. The standard relation in quantum field theory $F=-\log Z$ gives the generating function for connected graphs.

The similarities between (\ref{discofever}) and (\ref{partu2}) suggest that there exists a matrix model reproducing the Penner quantities. This model should have arbitrary $k$-valent vertices and thus arbitrary $k$-powers. Additionally it has only a coupling constant, namely $t$. The quantity $\Phi(I, N)$ in (\ref{genfun}) can be expressed as an integration of the form
\be\lb{matrix}
\Phi(I, N)=\sum_{\sum (k-2)v_k=I}\frac{1}{2^{N/2} \pi^{N^2/2} \Pi_k v_k}\int_{H^{N\times N}}dM(H)\Pi_k \bigg(\frac{-{\mathrm{Tr}} M^k}{k}\bigg)^{v_k} e^{-\frac{1}{2}\; {\mathrm{Tr}} M^2},
\ee
and the sum is interpreted as a sum over the tuples $\{v_k\}^K_1$ satisfying the constraints $\sum (k-2)v_k=I$ and $v_1=v_2=0$. \footnote{From the physical point of view this last condition corresponds
to remove the tadpole and the self-energy insertions. } In these terms the partition function that reproduce (\ref{partu2}) is found to be
\be\lb{penner2}
Z(t, N)=\frac{ \int dM e^{-\frac{1}{t}\; \mathrm{Tr} \sum_{n=2}^{\infty}\frac{M^n(i\sqrt{t})^n}{n}}}{\int dM e^{-\frac{1}{2}\; \mathrm{Tr} M^2}},
\ee
or, equivalently
\be\lb{penner}
Z(t, N)=\frac{ \int dM e^{-\frac{1}{t}\; \mathrm{Tr}(i \sqrt{t}M+\log(I-\sqrt{t}M))}}{\int dM e^{-\frac{1}{2}\; \mathrm{Tr} M^2}}.
\ee
The last is the partition function for the Penner matrix model \cite{Penner}. By parameterizing the matrices as $M=e^{\Phi}$ it is known that by a change of variables that this model reduce to the matrix Liouville model.

\subsection{The Euler characteristic and the string expansion}
The computation of the partition function (\ref{penner}) allows to find the explicit expression for $\Phi(N, t)$ by (\ref{partu}). By rewriting $\Phi(N, t)$ as in (\ref{genfun2}) the orbifold Euler characteristics $\chi(M_g^s)$ may extracted. Thus the explicit value of (\ref{penner}) is of enumerative interest. An standard procedure to find the value of a generic matrix model partition function, in particular (\ref{penner}), is by the use of monic polynomials $P_n(x)=x^n+\sum_{i=1}^{n-1}P_m^{(i)} x^{n-i}$ orthogonal with respect to the measure defining the model \cite{Itzykson}. This condition means that
\be\lb{ortogonal}
\int d\mu(x) P_n(x)P_m(x)=\delta_{mn}h_n,
\ee
with $h_n$ being interpreted as the norm of the polynomial $P_n(x)$. When $h_n=1$ the polynomial basis $P_n(x)$ is called orthonormal. In situations in which the moments $\int_{-\infty}^{\infty} x^n d\mu(x)$ exist for every $n$, the monic polynomials $P_n(x)$ do exist and are unique, and satisfy the following recurrence relation
\be\lb{everver}
x P_n(x)=P_{n+1}(x)+S_n P_n(x)+R_n  P_{n-1}(x), \qquad h_{n+1}=R_{n+1} h_{n},
\ee
for some $x$-independent terms $S_n$ and $R_n$. The measure of the Penner model (\ref{penner}) is obtained by going
to the eigenvalue representation. Hermitian matrices are always diagonalizable and one can go to the eigenvalue representation $\lambda_i$ for which the partition function $Z(t,N)$ is represented as
\be\lb{partu3}
Z(t,N)=\frac{1}{2^{N/2} \Pi^{N^2/2} \Pi_1^N p!}\int_{R^N}\Pi_{i\neq j}(\lambda_i-\lambda_j)^2\Pi_{i=1}^N d\mu_t(\lambda_i),
\ee
where $d\mu_t(\lambda_i)$ denote the measure defined by the jacobian of the transformation $U M U^{-1}$ which takes $M$ to the diagonal form namely
\be\lb{mesu}
d\mu_i(\lambda)=e^{-\sum_{k\geq0}\frac{\lambda^{k+2}(i\sqrt{t})^k}{k+2}} d\lambda=e^{\frac{1}{t}(\log\frac{1}{(1-i\lambda \sqrt{t})}-i\lambda\sqrt{t})} d\lambda.
\ee
The last formula can be rewritten as
\be\lb{nesum}
d\mu_t(x)=-i\sqrt{t} (e t)^{-t-1}(-z)^{-t-1}e^{-z}dz,\qquad z=\frac{1}{t}(i x \sqrt{t}-1).
\ee
This is the measure for the Penner model and in fact, it is precisely the integrand in the integral definition of the Gamma function $\Gamma(x)$. The orthogonal polynomials for this norm are related to the monic Laguerre polynomials $L_{n}^{-t-1}(z)$  by the following relation
\be\lb{normita}
P_{n,t}(x)=\frac{t^{\frac{n}{2}}}{i^n}(-1)^{n}n! L_{n}^{-t-1}(z),
\ee
and the norms $h_n$ corresponding to these polynomials are
\be\lb{nromo}
h_n=\frac{t^n}{i^{2n}}(n!)^2\int |L_n^{-t-1}(z)|^2 d\mu_t(z)=\frac{t^n}{(-1)^n}(n!)^2\frac{2\pi\sqrt{t}(et)^{-t-1}}{\Gamma(\frac{1}{t})\Gamma(1-\frac{1}{t})}\Gamma(n-\frac{1}{t}+1).
\ee
In terms of the above quantities the partition function (\ref{penner}) can be calculated explicitly, together with the euler characteristic
 $\chi(M_g^s)$. The jacobian of the transformation
$U M U^{-1}$ which brings $M$ to the diagonal form is
\be\lb{polunomial}
\Delta(x)=\prod_{i\neq j}(x_i-x_j)=\det\|x_i^{j+1}\|=\det\|P_{j+1}(x_i)\|,\qquad 1\leq i,j\leq N,
\ee
and this together with (\ref{penner}) gives the following expression for the partition function
\be\lb{pennerexp2}
Z(t, N)=\frac{N!}{(2\pi)^{N/2}\prod_{p=1}^N P!}\prod_{n=0}^{N_1}h_n,
\ee
or, by taking into account (\ref{nromo})
\be\lb{pennerexp3}
Z(t, N)=\bigg(\frac{2\pi\sqrt{t}(et)^{-t-1}}{\Gamma(\frac{1}{t})\Gamma(1-\frac{1}{t})}\bigg)^{N}\frac{N!}{\prod_{p=1}^N P!}\prod_{n=0}^{N_1}\frac{t^n}{(-1)^n}n!\Gamma(n-\frac{1}{t}+1).
\ee
The last is already an explicit expression, but it can be further simplified by use of the following identity concerning Gamma functions
$$
\Gamma(n-\frac{1}{t}+1)=\frac{(-1)^n}{t^n}\prod_{p=1}^n(1-p t)\Gamma(1-\frac{1}{t}),
$$
which allows to rewrite (\ref{pennerexp3}) as
\be\lb{pennerexp4}
Z(t,N)=\bigg(\frac{2\pi\sqrt{t}(et)^{-t-1}}{\Gamma(\frac{1}{t})}\bigg)^N\prod_{p=1}^{N-1}(1-p t)^{N-p}.
\ee
The generating function for connected graph $F(N,t)$ is the logarithm of (\ref{pennerexp4}) namely
\be\lb{pennerexp5}
F(t, N)=-N \mu(\frac{1}{t})+\sum_{p=1}^{N-1}(N-p)\log(1-p t).
\ee
Here we have defined the Planna function $\mu(\frac{1}{t})=\log \Gamma(\frac{1}{t})+(\frac{1}{t}-\frac{1}{2})\log t+\frac{1}{t}-\frac{1}{2}\log(2\pi)$ which has the series expansion
\be\lb{expansivo}
\mu(\frac{1}{t})=\sum_{m=1}^{\infty}\frac{B_{2m}}{2m(2m-1)}t^{2m-1},
\ee
with $B_n$ being the usual Bernoulli numbers. The last formula together with (\ref{pennerexp5}) gives the following expression for the free energy
\be\lb{pennerexp}
F(t, N)=N \sum_{m=1}^{\infty}\frac{B_{2m}}{2m(2m-1)}t^{2m-1}+\sum_{p=1}^{N-1}(N-p)\log(1-p t).
\ee
The Euler characteristic $\chi(M_g^s)$ is determined by expressing the last logarithm as an expansion in the indeterminate $N$ of the form (\ref{genfun2}). This can be achieved by expanding the logarithm in (\ref{pennerexp}) in powers of $t$ and by use of the combinatorial identity
$$
\sum_{p=1}^{N_1}p^v=\frac{1}{v+1}\bigg(\sum_{k=0}^{v+1}\bigg(\begin{array}{c}
                                                                                       v+1 \\
                                                                                       k
                                                                                     \end{array}\bigg) B_k N^{v+1-k}-B_{v+1}\bigg)
$$
\be\lb{portu}
=\frac{N^{v+1}}{v+1}-\frac{N^v}{2}+\sum_{k=1}^{[v/2]}\bigg(\begin{array}{c}
                                                                                       v \\
                                                                                       2k-1
                                                                                     \end{array}\bigg) \frac{B_{2k}}{2k} N^{v+1-2k},
\ee
with $[v/2]$ being the integer part of $v/2$. The application of these formulas gives
\be\lb{kids}
F(t,N)=\sum_{m=1}^{\infty}t^j\bigg[-\frac{N^{j+1}}{j(j+1)(j+2)}+\sum_{k=1}^{[(j+1)/2]}\frac{(2k-1)!(j-1)!}{(j+2-2k)!}\frac{B_{2k}}{2k!} N^{j+2-2k}\bigg].
\ee
Comparison between (\ref{kids}) and (\ref{genfun2}) shows that
the generating function $\Phi_c(2I, N)$ is given by
\be\lb{zievol}
\Phi_c(2I, N)=(-1)^I\bigg[-\frac{N^{j+1}}{j(j+1)(j+2)}+\sum_{k=1}^{[(j+1)/2]}\frac{(2k-1)!(j-1)!}{(j+2-2k)!}\frac{B_{2k}}{2k!} N^{j+2-2k}\bigg],
\ee
and taking into account the constraint $I=2g+s-2> 0$, $s\geq 1$ and $g\geq 0$ it may be seen that $k=g$. Then (\ref{genfun2}) shows
that the orbifold euler characteristic of the moduli space of punctured Riemann surfaces with $g\geq0$ and $s\geq 1$ is given by
\be\lb{erverp}
\chi(M_g^s)=(-1)^s\frac{(s+2g-3)!(2g-1)}{s!2g!}B_{2g}.
\ee
For $s=0$ and $g>1$ the result is
\be\lb{erverp2}
\chi(M_g^0)=\frac{B_{2g}}{4g(g-1)},
\ee
which is the orbifold characteristic of the moduli space without punctures.
It should be mentioned that there are alternative combinatorial expressions which can be used in order to calculate the Penner numbers, for example
\cite{Akhmedov}.

    An interesting feature of the Penner free energy is that the rescaling $t\to t/N$ converts it into
\be\lb{latino}
F(N,t)=-\sum_{I\geq1}\Phi(2I, N)\bigg(-\frac{t}{N}\bigg)^I=\sum_{g=0}^{\infty}F_g(t)N^{2-2g},
\ee
which is the topological expansion for the self-dual $c=1$ string found in \cite{Distler}. Here $$F_g(t)=\sum_{s\geq 1}(-1)^{s+1}\chi(M_g^s)t^{2g-2+s},$$ is the summation over the punctures, which leads to a closed string theory.

\subsection{A double scaling limit}
 Given a matrix model characterized by a coupling constant $t$ at large N, the perturbative genus expansion has a finite radius of convergence $t_c^g$ which may depend on the genus $g$ under consideration. But there are matrix models for which the critical points are independent of the genus, that is $t_c^g=t_c$ for all $g$. In this situation, when approaching the critical value, the average number of vertices in a typical Feynman diagram diverges and the diagram itself becomes a continuous two dimensional surface, and the free energy diverges as
\be\lb{diverge}
F_g(t)\sim (t-t_c)^{\chi(1+\frac{1}{2m})},\qquad \chi=2-2g,
\ee
with a constant $m$ which depends on the matrix model under consideration. In those cases one may define a limit, known as a double scaling limit
\be\lb{doubes}
N(t-t_c)^{\chi}=\mu,\qquad N\to \infty,\qquad t\to t_c,
\ee
with the parameter $\mu$ is held finite after the limit is taken. Note that the theory before the double scaling limit is defined in terms of two parameters
$N$ and $t$ and after the limit only one appears, namely $\mu$.

     The Penner model is one of those models admitting double scaling limit. This may be seen by expanding the logarithmic terms in (\ref{pennerexp}) by use of the Euler-Maclaurin formula
$$
\sum_{p=1}^{N-1}(N-p)\log(1-p t)=\int_1^N (N-x)\log(1-\frac{xt}{N})dx-\frac{\Delta f(N)}{2}+\frac{B_2}{2}\Delta f'(N)+...+\frac{B_{2k}}{(2k)!}\Delta f^{2k-1}(N),
$$
with
$$
\Delta f^{2k-1}(N)= f^{2k-1}(N)- f^{2k-1}(1).
$$
The integral in the last expression can be evaluated explicitly as
\be\lb{omrter}
\int_1^N (N-x)\log(1-\frac{xt}{N})dx=\frac{N^2}{2 t}(1-t)^2\log(1-t)-\frac{3N^2}{4}+\frac{N^2}{2t}.
\ee
Additionally the other terms in the series are
$$
\frac{f(N)}{2}=0,
$$
\be\lb{terma}
\frac{B_2}{2}f'(N)=-\frac{1}{12}\log(1-t),
\ee
$$
\frac{B_{2k}}{(2k)!}f^{2k-1}(N)=\frac{B_{2k}}{4k(k-1)}[N(1-t)]^{2-2k}t^{2k-2}.
$$
It is seen that all the terms of the expansion are divergent when $t\to 1$.
The continuum limit of the free energy is obtained by letting $N\to \infty$ and $t\to 1$ such that the quantity
$\mu=N(1-t)$ is fixed
\be\lb{conti}
F(\mu)=-\frac{\mu^2}{2} \log \mu+\frac{1}{12}\log \mu-\sum_{g\geq 2}\chi(M^0_g)\mu^{2-2g}.
\ee
This is the double scaling limit of the theory, and is the continuum description of the Penner model. It follows from (\ref{conti}) that the double
scaling of the Penner model is the generating function for the orbifold Euler characteristics of the moduli spaces of Riemann surfaces without punctures.

\subsection{Relation to CS with gauge group $SU(N)$ on $S^3$}

     One of the most interesting features of the Penner model (\ref{pennerexp}) gives direct information about the perturbative
and non perturbative part of the $SU(N)$ Chern-Simons on the sphere $S^3$. These remarkable properties follow partially from the higher
rank duality that is present in this CS theory, as it will be described below. The lagrangian of the $SU(N)$ CS theory on a three manifold $M$ is given by
$$
S=\frac{k}{4\pi}\int_{S^3} {\mathrm{Tr}}(dA\wedge A+\frac{2}{3}A\wedge A\wedge A),
$$
where A is a $SU(N)$ gauge connection on the trivial bundle
over $M$ and $k$ is the coupling constant. The explicit value for the partition function corresponding to this action when $M=S^3$ is
\be\lb{tore}
Z_{CS}^p(S^3,SU(N), k)=(k+N)^{-\frac{N}{2}}\sqrt{\frac{k+N}{N}}\prod_{j=1}^{N-1}\{2\sin\frac{j\pi}{N+k}\}^{N-j}.
\ee
It is customary to define the CS coupling constants
\be\lb{cs}
g_{cs}=\frac{2\pi N}{k+N},\qquad \lambda_{cs}=\frac{g_{cs}}{N}=\frac{2\pi}{k+N},
\ee
the first of them has a range between $0$ and $2\pi$.

The CS partition function introduced above possess the rank level duality \cite{Zemba}. This means that the partition function
is, up to a multiplicative factor, unchanged under the interchange of $k\leftrightarrow N$. This multiplicative factor can be read from
\be\lb{hk}
\frac{Z_{CS}^p(S^3,SU(N), k)}{Z_{CS}^p(S^3,SU(k), N)}=\sqrt{\frac{k}{N}}.
\ee
This symmetry also holds for the expectation values for the unknot and, for more general knots, the skein relations shows that the expectation
value of a knot in a representation $R$ of $SU(N)$ is the same as certain representation $\widetilde{R}$ in $S(k)$ defined in \cite{Naculich}.
The free energy of the model is the logarithm of (\ref{tore}) namely
\be\lb{shes}
F_{CS}^p(S^3,U(N))=-\log Z_{CS}^p(S^3,SU(N))=-\frac{N}{2}\log(k+N)+\sum_{j=1}^{N-1}(N-j)\log(2\sin\frac{\pi j}{k+n}).
\ee
The last expression may be decomposed as the sum of a perturbative and non perturbative piece by the trigonometric identity
$$
\sin \pi z=(\pi z)\; \prod_{n=1}^{\infty}\bigg(1-\frac{z^2}{n^2}\bigg),
$$
which gives the result
\be\lb{per}
F^p_{CS}(S^3,U(N))=\sum_{j=1}^{N-1}(N-j)\sum_{n=1}^{\infty}\log(1-\frac{j^2 g_{cs}^2}{4\pi^2 n^2 N^2}),
\ee
\be\lb{nonper}
F^{np}_{CS}(S^3,U(N))=\frac{N^2}{2}\log(k+N)-\frac{N(N-1)}{2}\log(2\pi)+\sum_{j=1}^{N-1}(N-j)\log j.
\ee
The perturbative part of CS (\ref{per}) is related to the Penner model as follows. Perform the variable change $t\to t/N$ in the Penner free energy (\ref{pennerexp}) and consider
two Penner models with opposite coupling constants $t$ and $-t$. When the free energy of both models are summed, then the terms with Bernoulli numbers cancel and the logarithmic terms survive. The sum is given by
$$
F(t,N)+F(-t,N)= \sum_{p=1}^{N-1}(N-j)\log(1-\frac{p^2 t^2}{N^2}).
$$
By introducing a succession of coupling constants $t_n$ parameterized as $t_n=g_{cs}/2\pi n$ and summing the last expression over $n$ it is obtained
$$
\sum_{n=1}^{\infty}F(t_n,N)+F(-t_n,N)=\sum_{j=1}^{N-1}(N-j)\sum_{n=1}^{\infty}\log(1-\frac{j^2 g_{cs}^2}{4\pi^2 n^2 N^2}).
$$
By comparing the last expression with (\ref{per}) it follows immediately that
\be\lb{cs}
F^{p}_{CS}(S^3,U(N))=\sum_{n=1}^{\infty}(F(N, t_n)+F(N,-t_n)),
\ee
i.e, the sum is equal to the perturbative part of the CS partition function (\ref{per}) \cite{Chair}. In the following we denote $t=t_1=g_{cs}/2\pi$, as this is the parameter which will be identified with the Penner one in (\ref{latino}). In fact, the relation with the Penner model shows that a double scaling limit exist when $t\to 1$.

 The relation between the Penner model and the $U(N)$ CS theory on $S^3$ is more deep than the one shown above. The $U(N)$ CS
theory on $S^3$ is also a model admitting a double scaling limit. Interestingly enough, the non perturbative part of CS (\ref{nonper}) can be obtained from the perturbative part (\ref{per}) by taking this double scaling limit. This result is a consequence of the rank level duality signaled in (\ref{hk}). Moreover, the limit $t\to 1$ is equivalent to $g_{cs}\to 2\pi$. The rank level duality $k\leftrightarrow N$ interchanges $g_{cs}=2\pi$ with $g_{cs}=0$. Therefore one may expect to read non perturbative contributions around $g_{cs}=0$ by a double scaling limit around $g_{cs}=2\pi$.

The above assertions can be made explicit as follows. The expression (\ref{per}) may be
expanded by using the definition of the $\zeta$ function \be\lb{eta}\sum_{n=1}^{\infty} n^{-2k}=\zeta(2k),\ee  together with the binomial series (\ref{portu}). The final result is
\be\lb{ixpans}
F_{CS}^{p}=\sum_{g=0}^{\infty}\sum_{h=2}^{\infty}F^p_{g,h} \lambda_{cs}^{2g-2+h}N^h,
\ee
with
$$
F^{p}_{0,h}=-\frac{2\zeta(h-2)}{(2\pi)^{h-2}(h-2)(h-1)h},
$$
\be\lb{uxpans}
F^{p}_{1,h}=\frac{\zeta(h)}{6(2\pi)^{h}h},
\ee
$$
F^{p}_{g,h}=-\frac{2\zeta(2g-2+h)}{(2\pi)^{2g-2+h}}\bigg(\begin{array}{c}
                                                                                      2g-3+h \\
                                                                                       h
                                                                                     \end{array}\bigg)\frac{B_{2g}}{2g(2g-2)}.
$$
Additionally, the non perturbative part (\ref{nonper}) can be expressed as
\be\lb{nonper}
F^{np}_{CS}(S^3,U(N))=\log\bigg(\frac{(2\pi\lambda)^{\frac{N^2}{2}} G_2(N+1)}{(2\pi)^{\frac{N(N+1)}{2}}}\bigg),
\ee
with $G_2(N+1)=1!2!..(N-1)!$. It follows that $G_2(N+1)$ is the Barnes function $G(z)$ evaluated on integer arguments, whose asymptotic form for large values of $z$ is known to be
$$
G(z+1)=\frac{1}{12}-\log A+\frac{z}{2}\log(2\pi)+\bigg(\frac{z^2}{2}-\frac{1}{12}\bigg)\log z-\frac{3z^2}{4}+\sum_{k=1}^{N}\frac{B_{2k+2}}{4k(k+1)z^{2k}}+O(\frac{1}{z^{2N+2}}),\qquad z>> 1.
$$
By use of this asymptotic expansion the following large $N$ limit for the non perturbative CS free energy is obtained
$$
F^{np}_{0}=\frac{N^2}{2}(\log(N\lambda_{cs})-\frac{3}{2}),
$$
\be\lb{optiv}
F^{np}_{1}=-\frac{1}{12}\log(N)+\zeta'(1),
\ee
$$
F^{np}_{g}=N^{2-2g}\chi(M^0_g).
$$
Besides, by defining the succession of parameters
\be\lb{qututu}
\nu_n=\frac{2\pi N}{g_{cs}}(\frac{g_{cs}}{2\pi}-n),
\ee
it follows from (\ref{cs}) that the double scale limit is when $N\to \infty$ and $g_{cs}\to 2\pi$ such that $\nu_1$ is constant. In this limit
the perturbative part of (\ref{cs}) becomes the Penner string expansion (\ref{conti}) by making the replacement $\mu_1\to\mu$.
Additionally, by comparing (\ref{optiv}) with the terms of the double scaled free energy of the Penner model (\ref{conti}) it is immediate that they are essentially the same, up to a redefinition $\mu\to N$. This is an interesting feature, and suggest that in certain cases one may read non perturbative contributions in field theory from the perturbative expansions. This issue was discussed in the context of geometrical engineering of gauge theories for instance in \cite{Gomez}.

\subsection{The closed string version}
The perturbative free energy (\ref{ixpans}) is an expansion in terms of open string amplitudes $F_{g,h}^p$ given in (\ref{uxpans}). But it may expressed as well as a summation over closed string amplitudes $F_{g}^p$ as
\be\lb{mxpans}
F_{CS}^{p}=\sum_{g=0}^{\infty}F^p_{g} \lambda_{cs}^{2g-2},
\ee
in which $F_{g}^p$ is defined as a sum over the holes $h$ of the Riemann surface
\be\lb{cloexp}
F_g^p(g_{cs})=\sum_{h=1}^{\infty}F^{p}_{g,h}g_{cs}^h.
\ee
This can be achieved with the definition of the zeta function (\ref{eta}) together with the binomial
identity
\be\lb{binin}
\frac{1}{(1-x)^q}=\sum_{n=0}^{\infty}\bigg(\begin{array}{c}
                                                                                       q+n-1 \\
                                                                                       n
                                                                                     \end{array}\bigg)x^n
\ee
gives that
$$
F^p_g(g_{cs})=\frac{(-1)^g |B_{2g}B_{2g-2}|}{(2g-2)!2g(2g-2)}-\frac{B_{2g}}{2g(2g-2)}(\frac{1}{g_{cs})^{2g-2}}
$$
\be\lb{klose}
+\frac{B_{2g}}{2g(2g-2)}\sum_{n=-\infty}^{\infty}\frac{1}{(g_{cs}+2\pi n)^{2g-2}}.
\ee
The non perturbative expansion (\ref{nonper}) can be expressed as
$$
F^{np}(g_{cs})=\sum_{g=0}F_g^{np}(g_{cs}) g_{s}^{2g-2},
$$
with
$$
F_g^{np}(g_{cs})=\frac{B_{2g}}{2g(2g-2)}\frac{1}{g_{cs}^{2g-2}},\qquad g\geq2.
$$
Note that this term cancels the second term in (\ref{klose}).
By using the identity
$$
\sum_{n=-\infty}^{\infty}\frac{1}{n+z}=\frac{2i\pi}{1-e^{2i\pi z}},
$$
and the definition of the polylogarithm
\be\lb{polilog}
Li_j(x)=\bigg(x\frac{d}{dx}\bigg)^{|j|}\frac{1}{1-x}=|j|!\frac{x^{|j|}}{(1-x)^{|j|+1}}+...
\ee
it is obtained finally
\be\lb{cerrada}
F_{g}(g_{cs})=\frac{(-1)^g |B_{2g}B_{2g-2}|}{(2g-2)!2g(2g-2)}+\frac{|B_{2g}|}{2g(2g-2)!}Li_{3-2g}(q),
\ee
valid for $g\geq 2$. Here the exponential $q=\exp{(-i g_{cs})}$ has been introduced for later convenience. In terms of the quantities (\ref{cerrada}) the free energy is expressed as a closed expansion of the form (\ref{mxpans}).

\section{A-model interpretation of the Abelian Nekrasov function}
\subsection{The A-model on the conifold and geometrical transitions}
  The CS theory with $U(N)$ gauge group on a compact three manifold $M$ admits an interpretation as an open string field
description for the topological string A-model on the manifold $T^{\ast}M$, which is a Calabi-Yau known as the cotangent bundle of $M$, for which $N$ branes wrap $M$, thus providing Dirichlet conditions
for the open strings. These resembles the situation for bosonic strings with Chan Paton factors $U(N)$, which are known to admit a description as a cubic non commutative field theory \cite{Wittu}, in which the string field is
a $U(N)$ matrix. For Calabi-Yau geometries of the form $T^{\ast}M$, the submanifold $M$ is lagrangian. Dirichlet conditions are assured by considering maps $X$ from the worldsheet to the background geometry which when restricted to the disconnected parts of the boundary $\partial \Sigma_g^h$  satisfy $X(\partial \Sigma_g^h)|_i\subset L_i$, with $L_i$ a lagrangian submanifold and $i$ denotes the $i$-th disconnected part of the boundary.

The simplest choice is $M=S^3$ for which the resulting CY manifold is $T^{\ast}S^3$. This situation corresponds to the CS theory with $U(N)$ gauge group over $S^3$ analogous to the one described in the previous section. The Calabi-Yau manifold $T^\ast S^3$ may be
characterized as follows. Consider the singular conifold, that is, the singular algebraic surface defined by
the points $(x,y,u,v)$ in $C^4$ satisfying the equation
\be \lb{sc}
xy - uv = 0.
\ee
This is a Calabi-Yau surface. It can be characterized as well as
\be\lb{determnin}
\det Z=\det\bigg(\begin{array}{cc}
 x & u \\
  y & v
\end{array}\bigg)=0.
 \ee
The torus action $T^2:(x,y,u,v)\to (e^{i\alpha}x,e^{-i\alpha}y,e^{i\beta}u,e^{-i\beta}v)$ leaves the equation (\ref{determnin}) invariant. The $\alpha$ and $\beta$ actions can be taken to generate the $(0,1)$ and $(1,0)$ cycles of $T^2$. Now, if a given point $(x,y,u,v)$ satisfies (\ref{sc}) then $\lambda(x,y,u,v)$ also satisfies it, thus (\ref{sc}) is a cone and is singular at the origin.
The geometry is a cone over $S^2\times S^3$. It is sometimes convenient to parameterize
in radial and angular coordinates as
$$
x=a_1a_2=r^{\frac{3}{2}}\cos\frac{\theta_1}{2} \cos\frac{\theta_2}{2}e^{i(\psi+\phi_1-\phi_2)},\qquad y=a_2a_4=r^{\frac{3}{2}}\cos\frac{\theta_1}{2} \sin\frac{\theta_2}{2}e^{i(\psi+\phi_1+\phi_2)},
$$
\be\lb{durv}
 u=a_2a_3=r^{\frac{3}{2}}\sin\frac{\theta_1}{2} \cos\frac{\theta_2}{2}e^{i(\psi-\phi_1-\phi_2)},\qquad v=a_1a_4=r^{\frac{3}{2}}\sin\frac{\theta_1}{2} \sin\frac{\theta_2}{2}e^{i(\psi-\phi_1+\phi_2)}.
\ee
 The canonical metric on this manifold is invariant under $SU(2)\times SU(2)\times U(1)$ since the transformation $L Z R^T$ and $U(1):(a_1, a_2,a_3, a_4)\to (e^{i\alpha}a_1, e^{i\alpha}a_2, e^{-i\alpha}a_3, e^{-i\alpha}a_4)$ leave the equation (\ref{determnin}) invariant.
There is a redundancy in the definition of the coordinates $a_i$ since given any set of them the redefinition $a_i\to \lambda a_i$ for $i=1,2$ and $a_i\to \lambda^{-1} a_i$ for $i=3,4$ gives valid coordinates as well. This redundancy can be eliminated by imposing the constraint
\be\lb{chas}
|a_1|^2+|a_2|^2-|a_3|^2-|a_4|^2=0,
\ee
quotiented with the $U(1)$ action described above.

The singular conifold admit two types of
Calabi-Yau desingularizations. The first is the \emph{deformed} conifold
which is obtained by introducing a complex parameter $\mu$ in (\ref{sc}) such that the limit $\mu\to 0$ gives the singular geometry
\be \lb{dc}
xy - uv = \mu.
\ee
This deformed version is still a Calabi-Yau surface but the conical singularity in the origin has been removed, in other words
the tip of the cone has been smoothed out. By a coordinate change
\be\lb{cws}
x=\eta_1+i\eta_2,\qquad y=\eta_1-i\eta_2,\qquad u=i\eta_3-\eta_4,\qquad v=i\eta_3+\eta_4,
\ee
with new complex coordinates $\eta_i=x_i+i p_i$ the algebraic equation (\ref{dc}) can be expressed as
\be\lb{misyu}
\sum_{i=1}^4(x_i^2-p_i^2)=\mu,\qquad \sum_{i=1}^4x_i p_i=0.
\ee
The last expression shows explicitly the $S^3$ inside the geometry, which is given
 by the locus where all $p_i=0$. The second equation (\ref{misyu}) shows that $p_i$
are the coordinates of the cotangent bundle. The full geometry where we include also the
imaginary parts of $x_i$ is in fact diffeomorphic to the cotangent bundle $T^*S^3$.
At infinity the conifold looks like $S^2 \times S^3$.  As we move from
infinity toward the origin both $S^2$ and $S^3$ shrink,
until the $S^2$ disappears and the $S^3$ is left with radius $\mu$.

   There exist still another CY resolution of (\ref{sc}), topologically inequivalent to the deformed conifold, known as the \emph{resolved} conifold.
The equation (\ref{sc}) is equivalent to the existence of nontrivial solutions of the linear system
\be\lb{equno}
u z=x,\qquad v z'=y,
\end{equation}
with the new coordinates $z$ and $z'$ parameterize $CP(1)\simeq S^2$ and are related by $z'=z^{-1}$.
Away from $(x,y,u,v) = (0,0,0,0)$ the
matrix has rank $1$, so $(z, z')$ solving are unique up to
an overall scaling and give no new coordinates for the manifold. But at the singular point $(x,y,u,v) = (0,0,0,0)$
any pair $(z, z')$  solve the equation.  Therefore this modified space is the same as the singular conifold, except that the point $(x,y,u,v)=(0,0,0,0)$ has been replaced by a whole $S^2$. This space is characterized as
\be\lb{resolvi}
\{(x,y,z,u, [z_1: z_2])\in C^4\times CP(1)\;| \;xu-yv=0,\; u z=x,\; v z'=y\},\qquad z=\frac{z_1}{z_2},\qquad z'=\frac{z_2}{z_1}
\ee
In terms of the coordinates $a_i$ introduced above the geometry can be reexpressed as
\be\lb{madrugue}
|a_1|^2+|a_2|^2-|a_3|^2-|a_4|^2=\mathrm{Re}(t),
\ee
quotiented with the $U(1)$ action  $U(1):(a_1, a_2,a_3, a_4)\to (e^{i\alpha}a_1, e^{i\alpha}a_2, e^{-i\alpha}a_3, e^{-i\alpha}a_4)$. The surface $a_3=a_4=0$ describes an sphere $S^2$ with coordinates $(a_1, a_2)$. The coordinates $a_3$ and $a_4$ are the fiber coordinates. The surface (\ref{madrugue}) is the resolved conifold $X^{t}$, and is a fibration of the form $X^{t}={\cal O}(-1)\oplus {\cal O}(-1)\rightarrow P^1$.

The deformed conifold $T^\ast S^3$ and the resolved conifold $X^{t}={\cal O}(-1)\oplus {\cal O}(-1)\rightarrow P^1$ are in some sense dual geometries, connected by a large N transition in the topological A model \cite{GV}. The branes of the A model on $T^\ast S^3$ are supported over the holes $h$ of the Riemann surface. But, as we discussed above, the A-model on $T^\ast S^3$ with $N$ branes wrapping the $S^3$ special lagrangian submanifold is equivalent to CS theory on $S^3$ and gauge group $U(N)$. After taking the double scaling limit one has the open perturbative expansion (\ref{ixpans})-(\ref{uxpans}). But this open expansion is equivalent to (\ref{cerrada}), for which the holes has disappeared together with the D-branes supported over them. It is natural to ask if there is a closed string theory whose free energy is (\ref{cerrada}).  If this closed string version do exist, then the absence of the D-branes should be compensated by a change in the background geometry. The Gopakumar and Vafa conjecture states that this closed string theory do exist, and is the topological closed string on the resolved conifold $X^{t}={\cal O}(-1)\oplus {\cal O}(-1)\rightarrow P^1$ \cite{GV}. In other words, CS theory with $U(N)$ on $S^3$ at large $N$ should be equivalent to the topological string on the
resolved conifold. The free energy for this topological string is given by
\be\lb{conref}
F_{r}(g_s,t)=\sum_{n=1}^{\infty}\frac{q^n}{n (2 \sin\frac{n g_s}{2})^2},
\ee
with $q$ the parameter defined below (\ref{cerrada}). It was explicitly checked in \cite{GV}, by use of the mathematical identities introduced in the previous sections, that free energy (\ref{conref}) coincides with (\ref{cerrada}) under the identification of the string coupling constant $g_{cs}$ and the CS one given by \be\lb{odk}g_{s}=i g_{cs}.\ee
This provides a partial consistency argument for this conjecture.

\subsection{Geometric transitions at the level of observables}

The correspondence described above between the free energy of topological strings on $T^\ast S^3$ at large $N$ and the resolved conifold $X^{t}={\cal O}(-1)\oplus {\cal O}(-1)\rightarrow P^1$ is supposed to hold also at the level of the observables \cite{Ooguvafa}-\cite{Ooguvafa2}. In the $U(N)$ CS theory on $S^3$ these observables are composed by Wilson loops operators, which are constructed in terms of a knot $K$, i.e, an embedded oriented circle in $S^3$, and in terms of a representation $R$ of $U(N)$ as follows
\be\lb{wl}
W_R={\mathrm{Tr}}_R P \exp\oint_{K} A.
\ee
Here $P$ denote the usual path ordering operation. A correspondence between branes in the A-model and the CS theory should be related to correspondence between lagrangian submanifolds and knots. Such correspondence is indeed known. For a given knot $K$ parameterized as $q(s)$ with $0\leq s\leq 2\pi$ in $S^3$ the following lagrangian submanifold $C_k$ in the Calabi-Yau $T^\ast S^3$ can be constructed \cite{Ooguvafa2}
\be\lb{waiting}
C_k=\{(q(s), p)\in T^{\ast}S^3| \sum_i \dot{q}_i p_i=0\}.
\ee
 Here $p_i$ are the coordinates of the cotangent bundle introduced in (\ref{misyu}). The 3-folds (\ref{waiting}) have the topology of a solid tori $R^1\times S^2$ and are known as conormal lagrangian submanifolds. These manifolds are known to be compact and to have a boundary at infinite, which is topologically a $T^2$ with homology basis $\alpha$ and $\beta$. The first cycle is the contractible one inside $C_k$, the second is not.

  When a geometrical transition from $T^{\ast}S^3$ to resolved conifold $X^{t}={\cal O}(-1)\oplus {\cal O}(-1)\rightarrow P^1$ takes place, the lagrangian submanifolds $C_k$ will evolve to a new geometry $\widetilde{C}_k$. It was argued in \cite{Gomis} that the resulting geometry is a lagrangian submanifold in the resolved conifold geometry of the Taubes type \cite{Taubes}. These lagrangian submanifolds are also constructed in terms of knots $K$, and have also the topology of $R^2\times S^1$ and with an asymptotic boundary $T^2$. In addition, they ends on the knot $K$ at the $S^3$ of the infinite of the resolved conifold. The authors of \cite{Gomis} postulated that the Taubes knots allows an holographic dual description of the CS theory. In this duality the bulk description is given by lagrangian D-branes ending in knots at the sphere at infinite, thus introducing sources for Wilson loop operators at the boundary.

The dual description mentioned above can be made more precise as follows. The lagrangian submanifold $C_k$ of $T^\ast S^3$ described in (\ref{waiting}) has the property that it intersects the $S^3$ along the knot $K$. When there are N branes wrapping the $S^3$ the topological A model gives the CS theory on $S^3$ with $U(N)$ gauge group. But in presence of M branes wrapping the 3-cycle $C_k$ there will appear strings stretched between both type of branes which corresponds to a massless complex field $\varphi$ transforming in the bifundamental $(N, \overline{M})$. This field is a fermion for branes and a boson for anti-branes. When integrated out by a one loop computation the resulting action for anti-branes is
\be\lb{modificado}
S_{M,N}(S^3)=S_{CS}(S^3, U(N))+S_{CS}(C_k, U(M))+\sum_{n=0}^\infty \frac{1}{n}{\mathrm{Tr}} U^n {\mathrm{Tr}} V^n,
\ee
with
\be\lb{wusu}
U={\mathrm{Tr}}_R P \exp\oint_{\alpha} A,\qquad V={\mathrm{Tr}}_R P \exp\oint_{\alpha} A',
\ee
$A$ and $A'$ being the $U(N)$ and $U(M)$ gauge connection respectively and the path integral is taken around the non contractible knot $\alpha$.
The exponential of the third operator in (\ref{modificado})
\be\lb{susu}
Z(U,V)=\exp(\sum_{n=0}^\infty \frac{1}{n}{\mathrm{Tr}} U^n {\mathrm{Tr}} V^n),
\ee
is known as the Ooguri-Vafa operator.

There exist other expressions for the Ooguri-Vafa operator (\ref{susu}) in the literature. One is
\be\lb{modificado2}
Z(U,V)=1+\sum_{\overrightarrow{k}}\frac{1}{z_{\overrightarrow{k}}}\Upsilon_{\overrightarrow{k}}(U)\Upsilon_{\overrightarrow{k}}(V),
\ee
where we have defined $\Upsilon_{\overrightarrow{k}}(U)=\prod_{j=1}^{\infty}({\mathrm{Tr}}
 U^j)^{k_j}$ and $z_{\overrightarrow{k}}=\prod_{j\geq 1}k_j! j^{k_j}$
and the sum is performed in such a way that $\sum k_j=|\overrightarrow{k}|=k$. Alternatively, by application of the Frobenius formula, the action (\ref{modificado2}) becomes a sum over representations $Q$
\be\lb{modificado3}
Z(U,V)=\sum_Q {\mathrm{Tr}}_Q U {\mathrm{Tr}}_Q V.
\ee
These traces appearing in the previous expression are sometimes known as the Schur functions $s_R(U)$ and the following relation holds
\be\lb{travc}
s_R(u_1,...,u_N)={\mathrm{Tr}}_{Q} U=\frac{\det u_j^{k_i+N-i}}{\det u_j^{N-i}},
\ee
with $(u_1,...,u_N)$  the eigenvalues of $U$.

For large $N$ the $M$ branes can be considered as probes, and by integrating
out the field $A$ an effective action will be obtained which is the sum $U(M)$ Chern-Simons plus quantum corrections $F(t, V)$. These corrections are by definition
\be\lb{qc}
\exp(-F(g_{cs},V))=\frac{1}{\int [DA]\exp(-S_{CS}(A, S^3))}\int[DA]\exp(-S_{CS}(S^3, A)+\sum_{n=0}^\infty \frac{1}{n}{\mathrm{Tr}} U^n {\mathrm{Tr}} V^n),
\ee
from where it follows that
\be\lb{qc2}
F(t, V)=-\log<Z(U,V)>_{S^3},\qquad Z(U, V)=\exp(\sum_{n=0}^\infty \frac{1}{n}{\mathrm{Tr}} U^n {\mathrm{Tr}} V^n).
\ee
The full effective action is then
\be\lb{fufu}
S_{eff}=S_{CS}(A', S^3)+F(g_{cs},V).
\ee
Since the resulting Lagrangian submanifold $C_k$ in the resolved geometry provides boundary
conditions for open strings, it was conjectured that the free energy of open topological strings with boundary
conditions specified by $C_k$ is identical to the free energy of the deformed Chern-Simons
theory with action (\ref{fufu}) \cite{Ooguvafa}-\cite{Ooguvafa2}. Taking into account this point of wiew it follows from the expressions (\ref{wusu}) and (\ref{qc2}) that
the effective action is
\be\lb{wufu2}
F(t, V)=-\log \bigg(e^{i S_{CS}(A)}\sum_Q  {\mathrm{Tr}}_Q P \exp\oint_{\alpha} A \int [DA'] e^{i S_{CS}(A')}{\mathrm{Tr}}_Q P \exp\oint_{\alpha} A'\bigg)
\ee
The path integral over the connection $A'$ can be interpreted as a wavefunction $<\psi|$ whose values depends on the boundary conditions
imposed at the infinite of (\ref{waiting}), whose boundary is $T^2$. Besides the insertion of the Wilson loop related to $A'$ gives another state
$|Q>$ at the torus at the infinite. Therefore the quantum action (\ref{wufu2}) may be interpreted in terms of a sum of representations as
\be\lb{wufu3}
F(t, V)=-\log\bigg( e^{i S_{CS}(A)} \sum_Q  {\mathrm{Tr}}_Q P \exp\oint_{\alpha} A <\psi|R>\bigg).
\ee
The representations form an orthonormal basis, thus the choice $<\psi|=<R|$ projects $Q=R$. As the topological A model corresponds to deformations of the Kahler 2-form $\omega$, the addition of branes corresponds to changes in the periods of $\omega$. In fact, the presence of $A'$ modifies these integrals by adding a boundary term \cite{Aspinwall}
$$
x_i=\int_{\alpha=\partial D}A_i+\int_{D}\omega.
$$
For a generic representation $R$ one has for anti-branes that \cite{Elitzur}
\be\lb{elitzur}
\oint_{\beta} A_i= g_s(R_i-i+\frac{M}{2}+\frac{1}{2}),\qquad i=1,2,..,M\qquad \oint_{D} \omega=g_s \frac{M}{2},
\ee
with $M$ the number of anti-branes, $D$ is a disk with ends in $\beta$, and the analogous formula hold for branes by replacing $R_i$ with $R_i^T$
\cite{Gomis}.

 At large N, when the branes has been integrated out, the Ooguri-Vafa conjecture states that the new geometry is the resolved conifold. But the M branes wrapping $C_k$ are still present and becomes branes wrapping the cycle $\widetilde{C}_k$ which is obtained from $C_k$ after the transition. As was mentioned above, these new cycles are the Taubes ones \cite{Gomis}. The Wilson loop operator defined for the representation $R_i$ becomes the bulk object corresponding to M lagrangian D-branes in the resolved conifold. Additionally, the sum of the holonomies (\ref{elitzur}) which we denote $x_i$ turns out to be
\be\lb{mudu}
x_i=g_s(R_i-i+M+\frac{1}{2}).
\ee
This quantity, which is calculated for the deformed conifold, is interpreted by this duality argument as the moduli for the lagrangian D-branes on the resolved conifold \cite{Aspinwall},\cite{Gomis}.

\subsection{Relation to $N=2$ $U(1)$ gauge theory}
 The Nekrasov function \cite{Nekrasov} was introduced in order to deal with the complicated grow of the instanton measure of super Yang-Mills
 when the instanton number increases. It tackles the problem by embedding
the N=2 theory into a background called $\Omega$ background, which is a twist of the $R^4$ bundle characterized by two complex parameters
$\epsilon_1$, $\epsilon_2$. The partition function of this theory has the advantage that it can be analytically performed to arbitrary order in the instanton expansion. The instanton part of the Seiberg-Witten theory is obtained through the
Nekrasov partition function by the limit $\epsilon_1=-\epsilon_2=g_s\to 0$. The case $\epsilon_1=-\epsilon_2=g_s$ was found to be related to string theories and matrix models already in \cite{Huang}-\cite{Klemm}.

In the following the attention will be focused in the non commutative $N=2$ $U(1)$ gauge theory with $\epsilon_1=-\epsilon_2=g_s$, whose partition function is given \cite{Nekrasov}-\cite{Losev} as
\be\lb{nekrasov2}
Z_{inst}(a, \Lambda, g_s)=\sum_{\lambda}\frac{m_k^2}{g_s^{2|\lambda|}}\; {\mathrm{exp}}(-\frac{1}{g_s^2}\sum_{k\geq 1}t_k {\mathrm{ch}}_{k+1}(a,\lambda)),
\ee
where $t_k$ are coupling constants, $\lambda=(k_1,..., k_j,...)$ is a partition, ${\mathrm{ch}}_{k+1}(a,\lambda)$ are its Chern characters
and $a$ is the vacuum expectation value of the vector multiplets. The quantity $m_k$ is the Plancherel measure for the partition, and is defined as
\be\lb{hook}
m_{\lambda}=\prod_{u\in \lambda}\frac{1}{h(u)}=\frac{{\mathrm{dim}}R_k}{k!}=\prod_{i<j}\frac{k_i-k_j+j-i}{j-i},
\ee
with $u(i,j)$ being a box in the Young diagramm of $\lambda$. The small phase of the theory corresponds to the choice $t_k=0$ except for $k=1$. In this phase the Chern character is given by $ch_2(a,\lambda)=a^2+2g_s^2|\lambda|$ and the partition function reduce to
\be\lb{nurr}
Z_{inst}(a, \Lambda, g_s)=\sum_{\lambda}\frac{m^2_\lambda}{g_s^{2|\lambda|}}\; \mathrm{\exp}(-\frac{t_1}{g_s^2} a^2-2t_1|\lambda|).
\ee
The Plancherel measure defined in (\ref{hook}) can be expressed in integral forms in terms of a function
$\gamma_{g_s}(x)$ which satisfies the difference equation
\be\lb{dif}
\gamma_{g_s}(x+\epsilon)-\gamma_{g_s}(x-\epsilon)-2\gamma_{g_s}(x)=\log(x),
\ee
and in terms of the piecewise linear function $f_{\lambda}(x)$ defined by the upper part of the Young diagram corresponding to the partition $\lambda$, known as the profile function
\be\lb{fval}
f_{\lambda}(x)=|x-a|+\rho(x),
\ee
$$
\rho(x)=\sum_{i=1}^{\infty}|x-a+g_s k_i+g_s (i-1)|-|x-a+g_s k_i+g_s i|+|x-a+g_s i|+|x-a+g_s (i-1)|.
$$
From the last expression it may be shown that the second derivatives $f''_{\lambda}(x)$ contains only terms proportional to Dirac delta. In these terms it may be shown
that the Plancherel measure is
\be\lb{fucnti}
m_{\lambda}=\exp\bigg(-\frac{1}{8}\int_{x_1>x_2} dx_1 dx_2 f''_{\lambda}(x_1)f''_{\lambda}(x_2)\gamma_{g_s}(x_1-x_2,\Lambda)\bigg),
\ee
The kernel $\gamma_{g_s}(x,\Lambda)$ of the integral
(\ref{dif}) admits an integral expression
\be\lb{kernol}
\gamma_{g_s}(x,\Lambda)=\frac{d}{ds}\bigg(\frac{\Lambda^2}{\Gamma(s)}\int_0^{\infty}\frac{dt}{t^{1-s}}\frac{e^{-tx}}{(e^{g_s t}-1)(e^{-g_s t}-1)}\bigg)_{s=0},
\ee
and an expansion on $g_s$ given by
\be\lb{expan}
\gamma_{g_s}(x,\Lambda)=\sum_{g=0}^\infty g_s^{2-2g}\gamma_g(x,\Lambda)
\ee
$$
=\frac{1}{g_s^2}(\frac{1}{2}x^2\log\frac{x}{\Lambda}-\frac{3}{4}x^2)-\frac{1}{12}\log(\frac{x}{\Lambda})+\sum_{g=2}^{\infty}\frac{B_{2g}}{2g(2g-2)}(\frac{g_s}{x})^{2g-2}
$$
which is precisely the expansion of the $U(N)$ Chern-Simons on $S^3$ (\ref{optiv}) and (\ref{terma}).

It is convenient to introduce a deformed partition function which reduce to the Nekrasov one in the limit of small coupling constant $g_s$.
This is achieved by evaluating the Schur function (\ref{travc}) at the specific values $u_i=q^{-i+\frac{1}{2}}$
\be\lb{qd}
s_R(u_i=q^{-i+\frac{1}{2}})=q^{-\frac{1}{2}|R|}s_R(1,q,q^2,...,q^{N-1}),\qquad q=e^{-g_s}.
\ee
The resulting value for this function is proportional to the q-deformation of the dimension of $R$ namely
\be\lb{poas}
s_R(1,q,q^2,...,q^{N-1})=q^{n(\lambda)}\prod_{u\in\lambda}\frac{[n+c(u)]}{[h(u)]},
\ee
where $n(\lambda)=\sum_i k_i(i-1)$ and  $c(u)=i-j$ is the content of $u$ and $h(u)=k_i+k_j^t-i-j+1$ is the hook length for a given box $u=(i,j)$ of the diagram of $\lambda$. In addition, (\ref{poas}) is also related the knot invariant for the unknots in $S^3$, which is given by
\be\lb{unkot}
W_R(K)=s_R(u_i=q^{\frac{N-2i+1}{2}})=q^{N|R|}s_R(u_i=q^{-i+\frac{1}{2}})
\ee
The deformed partition function is defined in these terms as \footnote{This sort of partitions were considered in \cite{Szabo} in order to enumerate Donaldson-Thomas invariants
of local toric Calabi-Yau threefolds without compact divisors. }
\be\lb{dance}
\widetilde{Z}_{inst}(a, \Lambda, g_s)=\sum_{\lambda}s^2_\lambda(x_i=q^{i-\frac{1}{2}})\; {\mathrm{\exp}}(-\frac{t_1}{g_s^2} a^2-2t_1|\lambda|),
\ee
There exist an useful formula for the Schur function
\be\lb{usuv}
s_\lambda(x_i=q^{i-\frac{1}{2}})=q^{\frac{1}{4}\kappa(\lambda)}\sum_{(i,j)\in \lambda}\frac{1}{q^{h(i,j)}-q^{-h(i,j)}},
\ee
where $h(i, j)$ is the hook length of the box $(i, j)$ in the Young diagram and $\kappa(\lambda)=2\sum_{(i,j)\in \lambda}(j-i)$. For small $g_s$, that is, for $q\simeq 1$ it follows from (\ref{usuv}) that (\ref{dance}) reduce to the the partition function (\ref{nurr}) for the non commutative $U(1)$ gauge theory on $R^4$ with $N=2$ supersymmetry.

 The expression (\ref{dance}) appears in other contexts. For instance, it was considered in \cite{taquito}-\cite{taquito2} as a model of q-deformed random partitions corresponding to a five dimensional $U(1)$ gauge theory with a fifth dimension being a circle $S^1$. The limit $q\to 1$ is equivalent to tend the radius of the circle equal to zero.

\subsection{Small coupling limit of the abelian Nekrasov function and OV operators}

 In this section we will show that the partition function $\widetilde{Z}_{inst}(a, \Lambda, g_s)$ in (\ref{dance}) can be intrepreted as the mean value of the Ooguri-Vafa operator corresponding to the toric Calabi-Yau geometry considered in the references \cite{Bershadsky}. This CY geometry can be parameterized by the complex coordinates $(x,y, u, v)$ and is defined in algebraic form as
\be\lb{otrotro}
(z-\mu_1)(z-\mu_2)+uv=0,\qquad xy=z,\qquad \mu_1<0,\qquad 0<\mu_2.
\ee
As in the case of the singular conifold, there is a torus symmetry given by $T^2:(x,y,u,v)\to (e^{i\alpha}x,e^{-i\alpha}y,e^{i\beta}u,e^{-i\beta}v)$ which leaves the equation (\ref{otrotro}) invariant.
By a redefinition of coordinates the algebraic equation describing the geometry can be expressed as
\be\lb{otrotro2}
(x_1^2+x_2^2-R)^2+u_1^2+u_2^2-r^2=0, \qquad R=\frac{\mu_1+\mu_2}{2},\qquad r=\frac{\mu_1-\mu_2}{2},
\ee
which represents it as an ALE (asymptotically locally euclidean) fibration. The $\alpha$ action is degenerated at $z=0$. Similarly the $\beta$ cycle degenerates at the two branching points $z=\mu_i$. The cycles
over the intervals $\mu_1<z<0$ and $0<z<\mu_2$ are two three spheres $S^3$.  The surface $uv=\mu_1\mu_2$ connects the two $S^3$ as its boundaries
are two circles, every of them located in each sphere. The real part of the algebraic equation (\ref{otrotro2}) has the topology of $S^2\times S^1$. The parameter $r$ measures the radius of the $S^2$ and the radius of $S^1$ varies from $R+\sqrt{r}$ to $R-\sqrt{r}$. Alternatively, the parameter $r$ measures the separation of the two spheres located at $\mu_i$ and $R$
measures their distance to the origin. This manifold has $b_3=2$ and $b_{2}=1$, and $r$ is interpreted as the unique Kahler parameter.

As for the conifold, in the situation in which $N_i$ branes are wrapped on each cycle, an scalar $\varphi$ in the representation $(N_1, \overline{N}_2)$ will appear, corresponding to the bifundamental strings stretched between the two
sets of branes. The difference with the situation for the conifold (\ref{modificado}) is that
this complex scalar is
now massive, since the strings have a finite length, and its
mass is proportional to $r$. The resulting action is given by
\be\lb{modifica}
S_{M,N}(S^3)=S_{CS}(A_1)+S_{CS}(A_2)+\sum_{n=1}^\infty \frac{e^{-nr}}{n}{\mathrm{Tr}} U_1^n {\mathrm{Tr}} U_2^n,
\ee
where the operators $U_{1,2}$ are the holonomies of the gauge fields over the $S^1$ described above. The expression (\ref{modifica}) can alternatively be written as a sum of representations of the form \cite{Aganagic2}
\be\lb{modifica2}
S_{M,N}(S^3)=S_{CS}(A_1)+S_{CS}(A_2)+\sum_R{\mathrm{Tr}}_R U_1 e^{-l r} {\mathrm{Tr}}_R U_2.
\ee
with $l$ the number of boxes in the representation of $R$. In this situation the two boundaries of the annulus give two knots $K_1$ and $K_2$. This means that the total free energy is given by
\be\lb{tutus}
F=F_{CS}(N_1, g_s)+F_{CS}(N_2, g_s)+\log \sum_R e^{-lr}W_R(K_1) W_R(K_2).
\ee
These knots turn out to be two unknot, as was shown in \cite{Aganagic2} by use of a Heedegard decomposition of the two 3-spheres into solid tori.
The Wilson loop operators $W_R(K_i)$ is a function of the coupling constant $g_s=2\pi/(k_i+N_i)$ (which is the same for both CS theories) and the parameters $t_i=g_s N_i$. If the choice $N_1=N_2$ is made, then $t_1=t_2=g_s N_i$
and the Ooguri-Vafa operator corresponding to (\ref{tutus}) is
\be\lb{tutus2}
<Z(U,V)>=\sum_R e^{-lr}W_R(K_1) W_R(K_2)=\sum_{R}e^{-l r}W_{R}(K)W_{R}(K)=\sum_{R} M^{|R|}s^2_\lambda(x_i=q^{i-\frac{1}{2}}),
\ee
with $M=e^{-r}q^N$. This is exactly the same as (\ref{dance}) under the identification
\be\lb{udun}
{\mathrm{\exp}}(-2t_1)\longleftrightarrow M=e^{-r}q^{N},
\ee
up to an irrelevant factor ${\mathrm{\exp}}(-\frac{t_1}{g_s^2} a^2)$. Taking this into account, together with the discussion above (\ref{dance}) it follows that the abelian Nekrasov function (\ref{nurr}) is the small phase limit of the Ooguri-Vafa operator corresponding to the toric geometry (\ref{otrotro}) with $t_1=t_2$.

Under a geometrical transition, the parameters $t_1$ and $t_2$ become the Kahler parameters of the new geometry. There is a third Kahler parameter which is given by $2t=2r-t_1+t_2$ \cite{Grassi}. By taking the double scaling limit $N_i\to \infty$ and $g_s\to 0$ such that $t_i=g_s N_i$ are kept finite gives the following expression for the free energy
\be\lb{larmar}
F=-\log<Z(U, V)>=\sum_{d=1}^{\infty} \frac{e^{-d t_1}+e^{-dt_2}+e^{-dt}(1-e^{-dt_1})(1-e^{-dt_2})}{d(2\sin(d g_s/2))^2}.
\ee
The third term in (\ref{larmar}) is the contribution from the Ooguri-Vafa operator (\ref{tutus2}).

\section{ B-model interpretation of the Abelian Nekrasov function}

In the previous section we have identified the abelian Nekrasov function as the Ooguri-Vafa operator corresponding to the A model defined over some specific toric geometry. As this geometry admits a mirror it should be possible to give a direct B-model interpretation the abelian Nekrasov partition function. This description is explained in the present section.
\subsection{Topological B model as deformations of complex structures}
 The topological B-model with a generic Calabi-Yau geometry as a target space  corresponds to parameter deformations which affect the complex structure of the Calabi-Yau but leaving the Kahler structure intact. Several properties were exhaustively studied in \cite{KS}. For example, for the deformed conifold $T^\ast S^3$ a deformation affecting the complex structure $xy=\mu$ is
\be\lb{servve}
x(y+\sum_{n>0} n t_n x^{n-1})=\mu,
\ee
with $t_n$ certain parameters. It is known that the deformations affecting the complex are those which induce
a change in the periods of the holomorphic (3,0) form of the Calabi-Yau
\begin{eqnarray}\label{3-form}
\Omega =\frac{1}{4\pi^2}\frac{dx\wedge dy \wedge du}{u}.
\end{eqnarray}
In the topological B-model these deformations arise when there is one complex dimensional subspace inside the
Calabi-Yau with N B-branes wrapped over it. The inclusion of these B-branes induce the following change the periods of the holomorphic 3-form
\be\lb{shift}
\Delta \int_C \Omega=N g_s,
\ee
with C a 3-cycle linking the B-brane world-volume.

    The geometry of the deformed conifold is a particular "local" CY geometry. These local geometries are CY surfaces of the form
\be\lb{algeb}
uv=H(x,y),
\ee
with complex coordinates ($x$, $y$, $u$, $v$) and $H(x,y)$ a function on two of these variables. The resulting surface is generically a 3-dimensional complex one. For the deformed conifold $T^{\ast}S^3$ the function $H(x,y)$ can be read directly from the defining equation (\ref{dc}) and is $H(x,y)=xy-\mu$ with $\mu$ the deformation parameter. For any local geometry, the distinguished Riemann surface
\begin{equation}
\label{surface}
H(x,y)=0,
\end{equation}
is the locus where the fiber degenerate into two components $u=0$ and $v=0$. For any local Calabi-Yau and in particular for the deformed conifold, the degenerating locus of Riemann surface $H(x,y)=0$ generically has punctures. The deformation of the complex structure of the Calabi-Yau is achieved by placing non compact branes on these punctures.

The theory of \cite{KS} is described in terms of the so called Kodaira-Spencer fields. For local Calabi-Yau these fields can be reduced from 3 to 1 complex dimensions along the normal directions. The resulting fields are defined on the asymptotic regions $x\to \infty$ and $y\to \infty$ corresponding to the locus of the equation $H(x,y)=0$ as follows
\be\lb{conofold}
x=\partial_y\widetilde{\phi}(y),\qquad y=-\partial_x\phi(x).
\ee
The one form $\lambda=y dx=\partial \phi$ is the reduction of the complex $3$-form $\Omega$ integrated along the two normal directions.
The period of $\lambda$ integrated along a point $x_i$ on the Riemann surface where the brane intersects receive a non trivial contribution
\be\lb{nt}
\oint_{x_i} \lambda=\oint_{x_i} \partial \phi=g_s,
\ee
therefore branes in these punctures are affecting the complex structure.

The quantum mechanical operator for creating a brane $\psi(x_i)$ in the model is defined by following relation \cite{Aganagic3}
$$
<...\oint_{x_i} \partial\phi\; \psi(x_i)...>=g_s<...\psi(x_i)...>,
$$
which implies that $x$ corresponds to a disk amplitude operators
\be\lb{fino}
\psi(x)=\int^x y(x')dx'.
\ee
The creation of N branes located at diferent positions $x_i$ corresponds to the mean value
\be\lb{secua}
<N|\psi(x_1)....\psi(x_N)|W>,
\ee
where the state $<N|$ is an N-fermion state and $|W>$ is an state in $H$, the space of chiral bosons. The introduction
of these N branes correspond to a deformation of the complex structure as (\ref{servve}) and in fact the parameters $t_n$ in (\ref{servve}) are defined in terms of the position of the branes by the formula
\be\lb{p�rameter}
t_n=-\frac{g_s}{n}\sum_{i=1}^N x_i^{-n}.
\ee
Therefore the effect of placing these disk operators (or branes) is to deform the complex structure of the conifold as (\ref{servve}).

The variables $x$ and $y$ can be though as conjugated variables, in analogous way as the momentum $p$ and the coordinate $x$ in usual quantum mechanics. In fact the operator $\widetilde{\psi}(y)$ related to the $y$ patch in (\ref{conofold}) by
\be\lb{s}
\widetilde{\psi}(y)=\widehat{S}\psi(x)=\frac{1}{\sqrt{2\pi g_s}}\int dx e^{-\frac{xy}{g_s}}\psi(x),
\ee
which makes plausible this interpretation. In the last formula $\widehat{S}$ is playing a role as an S-matrix. The quantization of the complex moduli is realized by introducing quantum Kodaira-Spencer fields. The vacuum expectation value of the KS field in turn defines a
quantum Riemann surface which determines the CY with varied complex moduli.
In the quantization scheme one has the following commutation relations \cite{Hollands}
\begin{equation}
[y, x] =g_s.
\end{equation}
In the case of anti-branes, the action becomes the minus
of the action of branes, $S_{\bar B}=-S_B$, and thus, in the quantization, the conjugate
momentum of $x$ becomes $-y$ which results in the
commutation relation, $[y, x]=-g_s$.

  For multiple branes the relation (\ref{s}) is generalized to
\be\lb{fino}
<N|\widetilde{\psi}(y_1)....\widetilde{\psi}(y_N)|W>=\int (\prod_{j=1}^N dx_j)  e^{-\sum_{i=1}^N\frac{x_iy_i}{g_s}}<N|\psi(x_1)....\psi(x_N)|W>,
\ee
By use of the Harish-Chandra-Itzykson-Zuber formula it follows that these correlators can be expressed as a the partition function of the following matrix model
\be\lb{sara}
Z(t,\overline{t})=<N|\psi(x_1)....\psi(x_N)|W>=\int_{N\times N}dX (\det X)^{-N}e^{\frac{1}{g_s}{\mathrm{Tr}}(\sum_{n}t_n Tr X^n+X Y)},
\ee
with $\overline{t}_n=\frac{g_s}{n}{\mathrm{Tr}} Y^{-n}$. The  model (\ref{sara}) is known as the Imbimbo-Mukhi model \cite{Imimbo}.

\subsection{The B-model on mirror geometries}

The A-model on a generic Calabi-Yau manifold
$Y$ is equivalent to the B-model on the mirror Calabi-Yau, if this mirror exists. Two CY manifolds $Y$ and $\widetilde{Y}$ are mirror if they are topologically inequivalent and the $N=2$ A-sigma models with one of them as background is isomorphic to the B-sigma model with the other as background. It is well known that for any two mirror manifolds the Hodge numbers $h_{1,1}$ and $h_{2,1}$ are interchanged. This means that the deformations of the complex structure on one side are equivalent to Kahler deformations on the other side, and viceversa. As the A-model is related to Kahler deformation and the B-model is related to complex structure deformations this imply that the A-model on one of them, say $Y$, is equivalent to the B-model on the other one $\widetilde{Y}$, and viceversa.

The mirror dual of the resolved conifold was worked out explicitly in \cite{Hori}. To describe this mirror is convenient to introduce homogeneous coordinates $(y_1: y_2: y_3, y_4)=(e^{-Y_1}: e^{-Y_2}: e^{-Y_3}: e^{-Y_4})$ parameterizing CP(3). The Riemman surface describing the mirror geometry of the resolved conifold is
\be\lb{aju}
H(y_i)=y_1+y_2+y_3+y_4=0.
\ee
The constraint from the GNLS is that the coordinates $y_i$ describe a quadric in CP(3) given by \be\lb{quadve}y_1 y_2=e^{-t}y_3 y_4.\ee
These equations imply that
\be\lb{imply}
Y_1+Y_2-Y_3-Y_4=t,\qquad e^{-Y_1}+e^{-Y_2}+e^{-Y_3}+e^{-Y_4}=0.
\ee
The Riemann surface (\ref{aju}) has four punctures. At any of these punctures two of the $Y_i$ goes to infinite. The location of these punctures in CP(3) are $P_1=(1:0:-1:0)$, $P_2=(1:0:0:-1)$ $P_3=(0:1:0:-1)$ and $P_4=(0:1:-1:0)$. At these punctures one may choose a coordinate $y_i=1$ and express one of the remaining coordinates in terms of the other two by use of the quadric equation (\ref{quadve}). Denote the two remaining coordinates as $x$ and $y$. For $P_{1,3}$ the Riemann equation is
\be\lb{nico}
H(x,y)=1+e^{-x_{1,3}}+e^{y_{1,3}}+e^{y_{1,3}-x_{1,3}-t}=0,
\ee
while for $P_{2,4}$ it is instead
\be\lb{nico2}
H(x,y)=1+e^{-x_{2,4}}+e^{-y_{2,4}}+e^{-x_{2,4}-y_{2,4}-t}.
\ee
Note that $x$ and $p=y+i\pi$ are canonically conjugated coordinates, as $x\to \infty$ implies that $p\to 0$.
There is a remaining $SL(2, Z)$ freedom in choosing the coordinates, which corresponds to transformation which do not alter the periods of the coordinates $x$ and $y$
and the three form $\Omega$. By an particular $SL(2,Z)$ one can express the last equation as
\be\lb{nico2}
\widetilde{u}\widetilde{v}=H(\widetilde{x},\widetilde{y})=(e^{\widetilde{y}}-1)(e^{\widetilde{x}+\widetilde{y}}-1)-\mu,\qquad \mu=1-e^{-t}.
\ee
The parameter $\mu$ describe the complex structures on the manifold.

The Riemann surface arise (\ref{aju}) has a matrix model interpretation. The CS partition function for $U(N)$ gauge group over $S^3$ can be rewritten by use of the Weyl denominator formula as \cite{Tierz}, \cite{Aganagic}
\be\lb{matrux}
Z_{CS}=\int \prod_{i=1}^N du_i\prod_{i<j}(2\sinh \frac{u_i-u_j}{2})^2 e^{-\frac{1}{2 g_s}\sum_i u_i}.
\ee
By defining the new variables $m_i=e^{u_i}q^{-N}$ this can be reexpressed as \cite{Aganagic2}
\be\lb{mosad}
Z_{CS}=\int dM e^{-\frac{1}{g_s}{\mathrm{Tr}} (\log M)^2}.
\ee
This matrix model is in fact equivalent to the B-model on the mirror of the resolved conifold \cite{Aganagic2}. The spectral curve defining this model has been found in \cite{Houches} and it turns out to be given by the Riemann surface (\ref{nico}) defining the mirror of the resolved conifold. The density distribution $\rho(\lambda)$ of the eigenvalues of the model is calculated by the resolvent, which for the matrix model (\ref{mosad}) this is given by
\be\lb{resovle}
\omega_0(t)=\frac{1}{2t}\oint\frac{dz}{2\pi i}\frac{W'(z)}{p-z}\bigg(\frac{(p-a)(p-b)}{(z-a)(z-b)}\bigg)^{\frac{1}{2}}
\ee
with $W'(z)=\log z/z$. The solution satisfying the appropriate asymptotic is given by
\be\lb{resolvo}
\omega_0(t)=-\frac{1}{tp}\log\bigg(\frac{1+e^t p+\sqrt{(1+e^{-t}p)^2-4p}}{2p}\bigg),
\ee
and the density of eigenvalues corresponding to these model is
\be\lb{density}
\rho(\lambda)=\frac{1}{\pi t\lambda}\tan^{-1}\bigg(\frac{\sqrt{4\lambda-(1+e^{-t}\lambda)^2}}{1+e^t\lambda}\bigg).
\ee
By defining the function
$$
x(p)=t(1-p\omega_0(p))+i\pi
$$
and denoting $p=e^{t-y}$ it follows that $x$ satisfy the relation
\be\lb{b-model}
e^x+e^y+e^{y-x+t}+1=0,
\ee
which is the algebraic equation defining the Riemann surface defining the resolved conifold, namely (\ref{nico}).

\subsection{Multiple branes and geometric transitions}

The addition of non compact B-branes in the B-model (\ref{matrux}) is given by the introduction of the operator
\be\lb{branas}
\psi(x)=(-1)^N q^{N^2+\frac{1}{2}N}\det(e^{-x}-M),
\ee
which is equal for a single brane to
\be\lb{eqve}
<\psi(x)>=p_N(x).
\ee
The branes in the model are described in terms of the orthogonal polynomials corresponding for the measure of (\ref{mosad}). These are the Stieltjes-Wigert
polynomials given by \cite{Torso}
\be\lb{patron}
p_n(x)=(-1)^n q^{n^2+\frac{1}{2}n}\sum_{k=0}^n\bigg[\begin{array}{c}
                                                                                       n \\
                                                                                       k
                                                                                     \end{array}\bigg] q^{\frac{k(k-n)}{2}-k^2}(-q^{-\frac{1}{2}} x)^k,
\ee
where $q=e^{g_s}$ and the symbols appearing in this expression are given by
\be\lb{pasar}
\bigg[\begin{array}{c}
                                                                                       n \\
                                                                                       k
                                                                                     \end{array}\bigg] =\frac{[n]!}{[k]![n-k]!},
                                                                                     \qquad [n]!=q^{\frac{n}{2}}-q^{-\frac{n}{2}}.
\ee
These single correlation functions satisfy the Schrodinger equation corresponding to a Hamiltonian
obtained by replacing the function $H(x,y)$ describing the mirror of the resolved conifold with $\widehat{H}(x,-g_s\partial_x)$,
which in this case is given by
\be\lb{corrol}
\widehat{H}(x,-g_s\partial_x)<\psi(x)>
=(q^N- e^{-g_s\partial_x}-q^{-\frac{1}{2}} e^x e^{-g_s\partial_x}-q^{-\frac{1}{2}}e^x e^{-2g_s\partial_x})<\psi(x)>=0.
\ee
This is the equation for the spectral curve of the matrix model translated into the operator formalism \cite{Poor}-\cite{Yi}. Besides, the multi correlation functions for branes at different position is given by \cite{Morozov}
\be\lb{eqve}
<\prod_{i=1}^k B(x_i)>=\frac{\det (p_{N+j-1}(x_i))}{\Delta(x)}.
\ee
The insertion of branes in the matrix model can be implemented by adding the following term in the matrix model
\be\lb{adding}
Z_{B}=\int dM \mathrm{exp}\bigg(-\frac{1}{g_s} \mathrm{Tr} [(\log M)^2+\log(M\otimes I_{m\times m}-I_{N\times N}\otimes X)]\bigg),
\ee
with $X=(x_1,..,x_m)$ being a diagonal $U(m)$ matrix whose diagonal elements parameterize the moduli of the branes.

The brane insertions can be translated into the A-model crystal picture of \cite{Crystal}, as there exist a map $Q$ which maps B-model calculations into A-model ones \cite{Okuyama}. In this picture, a non-compact A-brane is represented by a half-line in $R^3$ and it creates a defect in the lattice. The presence of defect is represented by the insertion of a certain fermion operator $\Psi_D$, such that the amplitude of a single brane is \cite{Okuyama}
\be\lb{sbr}
Z_A(x)=<\Psi_D(x)>,
\ee
with $x$ the modulus of the brane. These A-brane has a
topology $S_1 \times R_2$ and its amplitude is
\be\lb{manaan}
Z_A(x)=\prod_{n=1}^N (1-x q^n)=\frac{L(x, q)}{L(x e^{-t}, q)}
\ee
with $t=g_s N$ and $L(x,q)$ is the polylogarithm given by
\be\lb{mandolina}
L(x, q)=\prod_{k=1}^{\infty} \bigg(1-x q^k\bigg)= \exp(\sum_{k=1}^{\infty}\frac{x^k}{x[x]}).
\ee
The generalization to multiple branes is described in \cite{Okuda} and \cite{okudo}. These branes are described by the mean value of the Wilson loop operator
for the unknot in the deformed conifold $T^\ast S^3$, which in the crystal picture is represented as
\be\lb{wlo}
<W_R>=M(q)^2\exp(\sum_{n=1}^{\infty}\frac{e^{nt}}{n[n]^2})\prod_{i<j}(1-e^{x_i-x_j})\prod_{i=1}^M\exp(\sum_{n=1}^{\infty}\frac{e^{nx_i}+e^{n(t-x_i)}}{n[n]}),
\ee
with $x_i$ given in (\ref{mudu}) and $t=g_s(N+M)$. Following \cite{Gomis} we consider the situation in which the moduli of the branes is parameterized as $x_i=\mathrm{exp}-g_s(l+m-i+\frac{1}{2})$. These branes live in a resolved conifold with Kahler parameter $t=g_s(N+l)$. When the
numbers $l$ and $m$ are large these branes trigger a geometric transition  to a new
CY manifold called bubbling CY manifold \cite{Gomis}. The Kahler moduli for this new geometry
 $t_1=g_s m$, $t_2=g_s l$ and $t=g_s (N-m)$. In fact by introducting the moduli $x_i=\mathrm{exp}-g_s(l+m-i+\frac{1}{2})$ into (\ref{wlo}) it is obtained the partition function
 \be\lb{explota}
<W_R>=M(q)^2\exp(\sum_{d=1}^{\infty} \frac{e^{-d t_1}+e^{-dt_2}+e^{-dt}(1-e^{-dt_1})(1-e^{-dt_2})}{d(2\sin(d g_s/2))^2}).
\ee
 In the situation in which the two Kahler moduli are equal the logarithm of this quantity coincides, up to an additive term, with the free energy (\ref{larmar}). This means that we have reproduced the large $N$ limit of the partition (\ref{tutus2}) whose small phase limit the abelian Nekrasov function (\ref{nurr}), in the B-model setup.
\\

{\bf Acknowledgement:} The author is grateful to G. Giribet, F. Cuckierman, M. Schvellinger for motivating discussions, to S. Kovacs for explanations
 about $N=2$ gauge theories some time ago and to J. Fernandez for some discussion about enumerative problems. This work is supported by CONICET (Argentina) and by the
ANPCyT grant PICT-2007-00849.

\end{document}